\documentstyle[11pt,psfig]{article}
\begin{document}

\centerline{\bf A $\sim 14$ DAYS STAR WITH TWO PHASE-LOCKED }
\vskip 0.5 truecm
\centerline{\bf MODES OF PULSATION IN THE EROS DATABASE}
\vskip 1.0 truecm
\indent { \sl in Variable stars and the astrophysical return of microlensing surveys, ed R. Ferlet, J.P.Maillard,
\'editions fronti\`eres.}
\vskip 1.0 truecm
authors : {J.P. BEAULIEU$^{1,2}$, R.BUCHLER$^{3}$,  M.J.GOUPIL$^{4}$, Z.KOLLATH$^{3,5}$,}
       $^{1}$ Kapteyn Laboratorium, Postbus 800, 9700 AV Groningen, The Netherlands. \\
       {$^{2}$ Institut d'Astrophysique de Paris, CNRS, 98bis Boulevard Arago, 
       F--75014 Paris, France. \\
        $^{3}$ Physics department, University of  Florida, Gainesville, FL 32611, USA.\\
       $^{4}$ DASGAL, Observatoire de Paris, Meudon 92195, France.\\
       $^{5}$ Konkoly observatory, Budapest, Hungary. \\
        }

\begin{abstract}{ 
 Using CCD photometry obtained by the EROS collaboration in 1991-1993, we have
discovered an LMC variable star with a light curve that is oscillating with a
mean period of $\sim 14$ days and an amplitude of $\sim$ 0.3 mag.  The oscillations
appear with irregular amplitude variations. 
 The Fourier spectrum shows that the pulsation of this star is phase locked between two modes of frequencies $f_0$ and
1.5$\times f_0$.  Moreover, this object has strong $H \alpha$ and $H \beta$
emission lines and neutral lines of Helium that suggest a spectral type between
late O and early B.  In a preliminary analysis, we derive a luminosity of $
L=3.4-3.8L_\odot$ and an effective temperature in the range $\log(T_{eff}) =3.85-4.2$.  }
 \end{abstract}

\section { Observations }

CCD photometry was obtained in a field of 0.5 square degree in the bar of the
LMC between 1991-1993 for EROS. About 2500 images spanning $\sim$130 days
were taken in two broad bandpass filters $B_E$ and $R_E$ centered respectively
on 490 and 670 nm in the 9192 campaign, and 5500 images  were taken of the
same field with a pair of very similar filters ($B_{E2},~R_{E2}$) for the
1992-1993 campaign.
We have systematically searched the EROS database for variable stars using the
modified periodogram technique\cite{MPG}\cite{UAI155} and the AoV
method\cite{SCW}.

Among the hundredth detected variable stars, we have
discovered a bright LMC variable star ($\alpha=5h18$m$10.5,~\delta=-69^035m59$, equinox 2000.0) 
with a 'period' of $\sim 14$ days and a particular behaviour, a
clear alternance between cycles with larger and smaller amplitudes (Fig 1).  
A Fourier fit with 8 independent frequencies leads to a spectrum  with two dominant
frequencies at $f_0$=825.69 pHz and at 1.5003$\times f_0$  (Fig.2).  This suggests that
the  pulsation of this star is phase locked between two modes of frequencies $f_0$ and
1.5$\times f_0$.

\begin{figure}
 \centerline{\psfig{figure=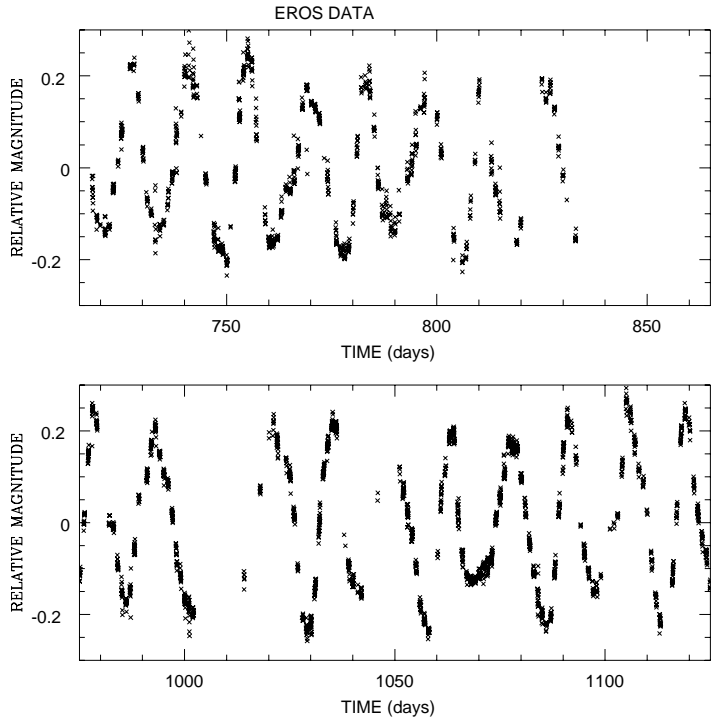,height=10cm,width=10cm}}
 \caption{ $B_E$ light curve from 1991-1992, and the begining of 1992-1993 campaign.}
\end{figure}

\begin{figure}
 \centerline{ \psfig{figure=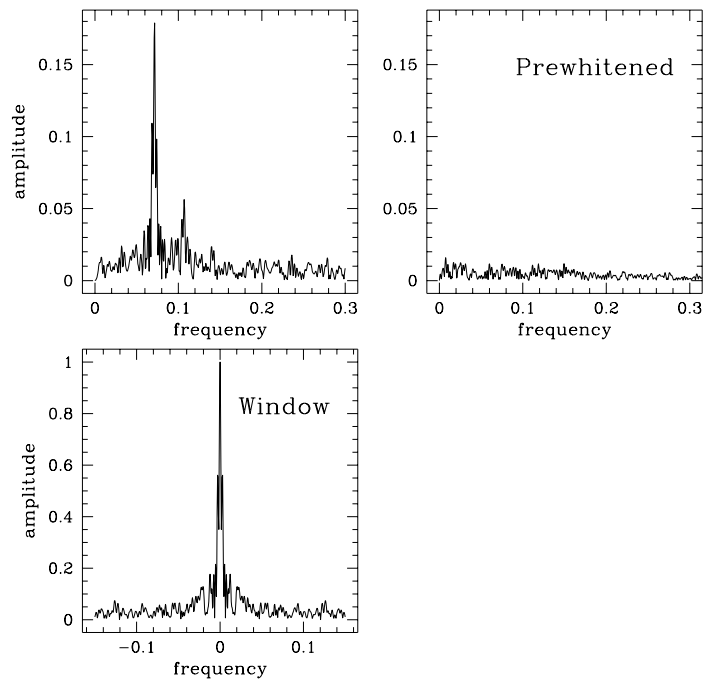,angle=0,height=10cm,width=10cm}}
 \caption{ Fourier spectrum of the 1991-1993 light curve,  of the prewhitened light curve in which 
the two frequencies $f_0$ and $f_1$ have been removed, and the spectral window. 
Notice the power around 0.07 (fundamental  frequency), and the secondary peak
around 0.11.}
\end{figure}

 From the photometry, assuming different values of reddening ( E(B-V) =0.10-0.30) 
and applying the temperature scale from Kurucz's atmospheric models,
we estimate the effective temperature of the star to be in the range
$\log(Teff) =3.85-4.2$. 

 The apparent magnitude of the star is $V_J=13.60$. 
Assuming a distance modulus to the LMC of $\mu_{LMC}=18.5$ mag, 
and a bolometric correction in the range $-0.3, -2$,  
we derive a luminosity in the range $ L=3.4-3.8L_\odot$.

Low resolution spectroscopy in the wavelength range $3700-7000 \AA$ has been
obtained at ESO la Silla in December 1995 with the ESO 1.5m equiped with a
Boller and Chivens spectrograph.  The resolution was $7 \AA$, and we get a signal 
to noise ratio of 50 at $6000 \AA$. This star  belongs to the LMC.
Its spectrum shows strong $H\alpha$  and $H \beta$ emission lines, 
and neutral lines of Helium indicating a spectral type between late O and early B (Fig 3).

 \begin{figure}
 \centerline{ \psfig{figure=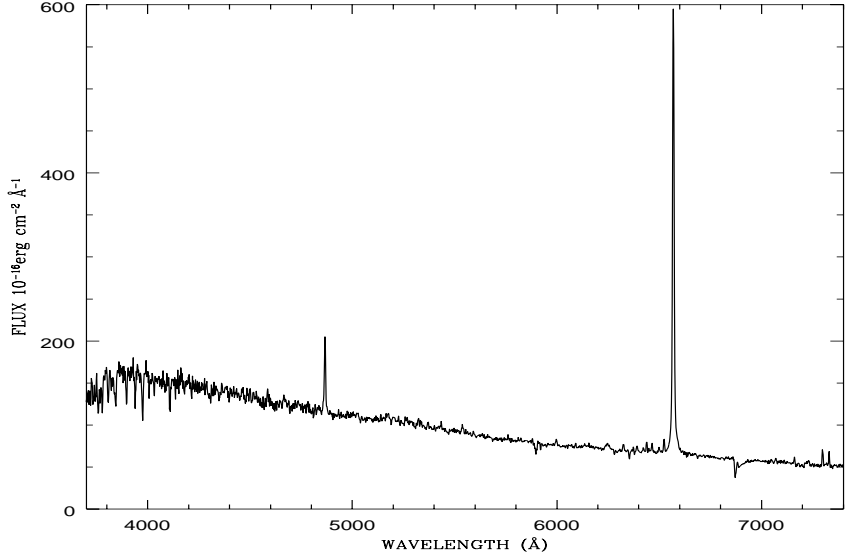,height=8cm,width=12cm}}
 \caption{Low resolution spectrum of the object.}
 \end{figure}


Further observational information are : \\
 \noindent (1) This star is located in the same area in the bar of the LMC,
 where 7 pre main sequence star candidates (PMSC) have been found\cite{PMS}.\\
 \noindent (2) It presents strong Balmer emission lines (equivalent width of $91 \AA$ for $H_\alpha$).\\
 \noindent (3) Its colour and spectrum suggest a spectral type between late O and early B.\\
 \noindent (4) We have no evidence for the presence of an extended $HII$ region around this object.\\
 \noindent (5) It is brighter than the PMSC which  we discovered in the LMC.\\
 \noindent (6) It presents an irregular photometric variability with a time
scale of $\approx 14$ days.  However, the photometric variability seems to be
due to a two mode phase-locked pulsation.  The observed large amplitude leads to
favour a radial pulsation rather a non-radial one.\\

Based on these arguments, we suggest that this star can be a pre main sequence object 
or a post-AGB star. A  companion poster \cite{ZOL} examine the possible nature of this object 
by means of a linear stability analysis of hydrostatic envelopes, and computations of  
hydrodynamical  models.

\long\def\jumpover#1{{}}

\jumpover{
 Which kind of mechanism
can produce this pulsation ? It can be due to the kappa mechanism, however the
up to date theoretical models show that for massive stars the instability to
different modes of pulsation is confined in a narrow region of the HR diagram
(the instability strip) where this star is not\cite{ZOL}. However, these
observational constraints can also lead to an interpretation as a low-mass
post-AGB star. The low mass star contain most of its mass in the tiny hot core
which is surrounded by an extended enveloppe with a virtually zero density
gradient. As a result, the pulsation in such a star will be strongly
non-adiabatic and non-linear. Moreover, luminous low-mass star are expected to
be unstable to different modes of pulsation through a wide range of
temperatures and luminosities in contrast to massive stars\cite{COX}.

Based on these arguments we favour the interpretation as a post-AGB star.  The
post-AGB phase (transition between AGB and white dwarf phase), is a very fast
evolution.  A strong confirmation of our interpretation would be the
observation of colour changes and period changes over years that would be
detectable with the complete EROS-EROSII curve and observations obtained by
MACHO in a few years.\\
}

acknowledgements : {This work has been supported by IFT at the University of
Florida, CNRS DASGAL, the IAP and NSF, and is based on observations held
at ESO La Silla.}

\vfill

\begin{thebibliography}{99}{\baselineskip 0.4cm
\bibitem{UAI155} Beaulieu J.P., 1995, in {\sl Astrophysical Application of Stellar Pulsation}, 
  Stobie R.S., Whitelock P.A. (eds.) ASP Conf. Ser. {\bf 83}, 260
\bibitem{PMS} Beaulieu J.P., Lamers H., et al., 1996,{\sl  Science } {\bf 272}, 995
\bibitem{COX} Cox J.P. et al., 1980, {\sl Space Science Reviews} {\bf 27}, 529.
\bibitem{MPG} Grison P., 1994, {\sl Astr. Astrophys.} {\bf 289}, 404
\bibitem{BIN95} Grison P., et al., 1995, {\sl Astr. Astrophys. Suppl. Ser.} {\bf 109}, 447
\bibitem{ZOL} Koll\`ath Z. et al.,  1997, this volume 
\bibitem{SCW} Schwarzenberg-Czerny, 1989  {\sl Mon. Not. R. astr. Soc} {\bf 241}, 153 
}
\end{thebibliography}
\end{document}